\documentclass[preprint,1p,11pt]{ISAS_IR}

\usepackage[english]{babel}
\usepackage[latin1]{inputenc}
\usepackage{lmodern}
\usepackage{amsmath}
\usepackage{amssymb}
\usepackage{epsfig}
\usepackage{epstopdf}

\def\Box#1{{\fboxsep2pt\fbox{$#1$}}}

\usepackage{theorem}
\theoremheaderfont{\bfseries}
\theoremstyle{plain}
{\theorembodyfont{\itshape} \newtheorem{Theorem}      {Theorem}   [section]}
{\theorembodyfont{\itshape} \newtheorem{Lemma}        {Lemma}     [section]}
{\theorembodyfont{\sffamily\small} }
{\theorembodyfont{\sffamily}\newtheorem{Corollary}    {Corollary} [section]}
{\theorembodyfont{\itshape} \newtheorem{Remark}       {Remark}    [section]}
{\theorembodyfont{\itshape} }

\newenvironment{Proof}
{\noindent{\scshape Proof.}}
{\hspace*{\fill}$\square$\bigskip\\\noindent}

\newlength\EqLen

\def\ScaleInner#1{%
\settowidth{\EqLen}{#1}
\ifdim\EqLen < \columnwidth%
\begin{equation*}%
\begin{minipage}{\EqLen}#1\end{minipage}%
\end{equation*}%
\else%
\begin{equation*}%
\resizebox{0.99\columnwidth}{!}{\begin{minipage}{\EqLen}#1\end{minipage}}%
\end{equation*}%
\fi%
}%

\def\Scale#1
  {
  \ScaleInner{$ #1 $} 
  }
\newcommand{\tr}{^{\rm T}}
\newcommand{\Ew}{{\rm E}}
\newcommand{\E}[2][\,]{\ensuremath{{\operatorfont{E}}_{{#1}}\!\!\left\{#2\right\}}}
\def\vec#1{\underline{#1}}
\def\mat#1{{\mathbf #1}}
\newcommand{\rvec}[1]{\ensuremath{\boldsymbol{\underline{#1}}}}

\newcommand{\estx}[2]{\ensuremath{\overline{\vec x}^{#1}_{#2}}}

\newcommand{\estCx}[2]{\ensuremath{\overline{\mat P}^{#1}_{#2}}}

\newcommand{\infx}[2]{\ensuremath{{\vec x}^{#1}_{#2}}}
\newcommand{\infxu}[2]{\ensuremath{{\vec x}^{u_{#1}}_{#2}}}
\newcommand{\infCx}[2]{\ensuremath{{\mat P}_{#2}}}
\newcommand{\infCz}[1]{\ensuremath{{\mat P}^{z}_{#1}}}
\newcommand{\infCw}[1]{\ensuremath{{\mat \Xi}^{#1}}}
\newcommand{\infCv}[2]{\ensuremath{{\mat \Theta}^{#1}}}
\newcommand{\infDelta}[2]{\ensuremath{\mathbf{\Delta}^{x_{#1}}_{#2}}}

\newcommand{\infG}[2]{\ensuremath{{\mat G}_{#2}}}
\newcommand{\infA}[2]{\ensuremath{{\mat A}_{#2}}}
\newcommand{\infB}[1]{\ensuremath{{\mat B}}}
\newcommand{\infK}[1]{\ensuremath{{\mat K}_{#1}}}
\newcommand{\infH}[2]{\ensuremath{{\mat C}^{#1}}}
\newcommand{\infL}[2]{\ensuremath{{\mat L}^{#1}_{#2}}}

\newcommand{\infxBackup}[2]{\ensuremath{{\vec x}^{#1'}_{#2}}}
\newcommand{\infxuBackup}[2]{\ensuremath{{\vec x}^{u_{#1'}}_{#2}}}
\newcommand{\infDeltaBackup}[2]{\ensuremath{\mathbf{\Delta}^{x_{#1'}}_{#2}}}

\newcommand{\truex}[1]{\ensuremath{{\rvec x}_{#1}}}
\newcommand{\truez}[2]{\ensuremath{{\vec y}^{#1}_{#2}}}
\newcommand{\trueu}[1]{\ensuremath{{\vec u}_{#1}}}
\newcommand{\truew}[1]{\ensuremath{{\rvec w}_{#1}}}
\newcommand{\truev}[2]{\ensuremath{{\rvec v}^{#1}_{#2}}}

\newcommand{\inv}[1]{\ensuremath{\big( {#1} \big)^{-1}}}
\newcommand{\transpose}[1]{\ensuremath{\big( {#1}\big)^\top}}

\begin{document}

\begin{frontmatter}
\title{Optimal Sequence-Based Control and Estimation of Networked Linear Systems}

\author[isas]{J\"org~Fischer}
\ead{joerg.fischer@kit.edu}
\author[isas]{Marc~Reinhardt}
\ead{marc.reinhardt@kit.edu}
\author[isas]{Uwe~D.~Hanebeck}
\ead{uwe.hanebeck@ieee.org}

\address[isas]{Intelligent Sensor-Actuator-Systems Laboratory (ISAS)\\
Institute for Anthropomatics\\
Karlsruhe Institute of Technology (KIT), Germany}

\begin{abstract}
In this paper, a unified approach to sequence-based control and estimation of linear networked systems with multiple sensors is proposed. Time delays and data losses in the controller-actuator-channel are compensated by sending sequences of control inputs. 
The sequence-based design paradigm is further extended to the sensor-controller-channels without increasing the load of the network. In this context, we present a recursive solution based on the Hypothesizing Distributed Kalman Filter (HKF) that is included in the overall sequence-based controller design.
\end{abstract}

\end{frontmatter}

\section{Introduction}
\label{sec:Introduction}
%
%
With advances in the development and distribution of modern network technologies such as Ethernet or WLAN (IEEE 802.11), general purpose networks for cordless communication have become an easy and cheap alternative to complex field buses. Although new flexibility is acquired, the communication is typically less reliable than in wired real-time architectures and, hence, users are faced with additional challenges regarding network-specific disturbances such as transmission delays and data losses. In control theory, these kinds of network-related problems are investigated in the area of Networked Control Systems~(NCS).

%
%
In the NCS community, several approaches have been proposed that consider transmission delays and stochastic data losses in the controller design (see \cite{Hespanha07} for an overview). In this contribution, we focus on a design philosophy called \textit{sequence-based control}, which stems from the idea that modern digital communication networks usually transmit data in form of atomic packets, which enforce that either all data of a packet is received or none. Therefore, the idea is to not only send the usual data over the network, but also extra information that is used to mitigate the network-induced effects. However, the transmission of extra information increases the network load and in general causes higher transmission delays and/or loss rates. 
Nevertheless, it is a common assumption that transmitting additional information within a data packet leads to an increased system performance as long as the additional information is not too extensive.

%
%
For the connection between controller and actuator (CA-channel), additional information is usually attached in form of predicted control inputs that are applicable at future time steps. The actuator stores these control inputs in a buffer so that it can fall back upon a predicted control input in case a data packet is lost or delayed.

%
%
In literature, three major lines of sequence-based controller designs can be distinguished. The first line of methods is based on a nominal controller that is designed for the system without consideration of networked-induced effects and that is extended afterwards to generate sequences~\cite{HeklerCDC12, liu2010predictive}. The second line of methods stems from Model Predictive Control~(MPC) theory, where the control sequences are generated as byproduct of solving an open-loop receding-horizon optimization problem~\cite{Gruene09}. In the third line of methods, the sequence-based NCS is formulated as a stochastic optimal control problem that is solved offline for some system classes~\cite{FischerArxiveACC12, gupta2006receding}.

In this contribution, we extend a result of the third line that has been derived in~\cite{FischerArxiveACC12} for linear systems and TCP-like network connections in two ways (see Sec.~\ref{sec:Problem_Formulation} for comments on TCP-like networks). First, we extend the system to multiple independent sensors that individually send partial state measurements over the network. We show that in this scenario the separation principle holds and discuss the structure of the optimal control law that depends on the minimum-mean-squared-error estimate of the state. Second, we apply the sequence-based design philosophy of the CA-channel to the data connection between sensors and controller (SE-channel).

%
%
An intuitive approach for the latter extension is that each sensor not only sends the most recent measurement to the controller, but also a (finite) set of measurements obtained in previous time steps~\cite{liu2010predictive}. This way, the estimation performance is improved since in case a sensor transmission has been dropped or delayed by the network, the missing information is also part of following transmissions. But, as pointed out above, the sent measurement sequences should not be too large. In particular, in the presence of multiple sensors, the network load increases considerably with the number of measurements transmitted per packet.

Hence, in this paper we investigate an approach how information contained in (possibly infinitely long) measurement sequences can be comprised in recursive variables. For that reason, we assume the sensors to be capable of performing minor processing tasks. While the Kalman Filter is the optimal estimator when all measurements can be processed at one node~\cite{Hespanha07}, the estimation quality is no longer optimal when local Kalman Filters are employed on multiple sensors~\cite{Chang1997}.

As we are interested in an estimation principle that is equivalent to processing the respective measurement sequences, we utilize recent results in estimation theory, where it has been shown in form of the so called Distributed Kalman Filter~(DKF)~\cite{Koch2008b, Govaers2011} that measurement data can be processed by a group of independent local sensors in a way that the fused result still yields a globally optimal result. 

Unfortunately, in the presence of lossy communication networks, the DKF does not work as it is necessary to have a complete set of all sensor data at the controller side to generate a state estimate. A generalization that manages to provide estimates even under uncertain conditions is the Hypothesizing Distributed Kalman Filter~(HKF)~\cite{Reinhardt2012a, Reinhardt2012c}. The second part of this contribution focuses on the extension and integration of the recursive HKF algorithm into the sequence-based controller design.

%
%
The following of the paper is structured as follows. After the problem formulation in Sec.~\ref{sec:Problem_Formulation}, we generalize the sequence-based control result from~\cite{FischerArxiveACC12} in Sec.~\ref{sec:Optimal_Controller_Design} by showing that the separation principle holds for further classes of available measurement information. The main part of this contribution is presented in Sec.~\ref{sec:Estimator_Design}, where we derive a generalization of the HKF in order to extend the sequence-based methodology to the measurement channel. In particular, we propose novel recursive sum formulas for the HKF, introduce a method to initialize estimates from measurements, and derive an extension of the HKF that is capable of handling the subsequent inclusion of control inputs at the controller side. After discussing the application of the proposed algorithm, we give a short summary and an outlook in~Sec.~\ref{Conclusion}.

%
%
\begin{figure}
\includegraphics[width=165mm]{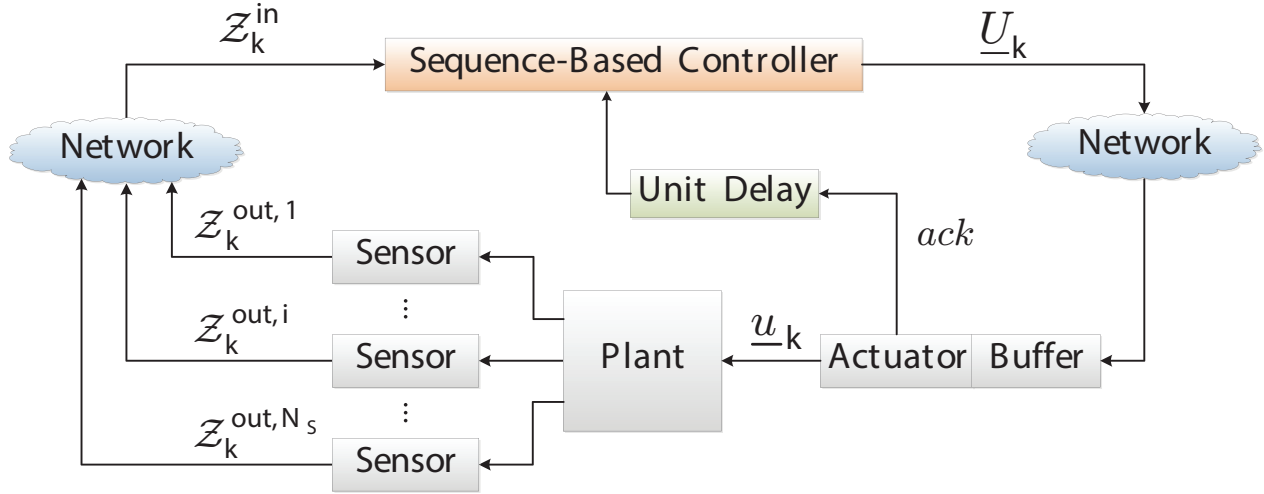}
\caption{The considered scenario with one actuator and multiple sensors that are connected to the controller via a network.}
\label{fig:system_overview}
\end{figure}

\section{Problem Formulation}
\label{sec:Problem_Formulation}
%
%
In this paper, we consider the system setup depicted in Fig.~\ref{fig:system_overview} consisting of a plant that is controlled and observed over a digital network. The plant is supposed to be linear and time-invariant and its output is not directly observable but measured by $M \in \mathbb{N}_{>0}$ independent sensors\footnote{The setup can easily be extended to the case of multiple actuators and a time-variant plant. In order to give a concise description of the unified sequence-based approach we limit ourselves to this simpler setup.} according to
\begin{align}
	\rvec{x}_{k+1} &= \mat{A} \rvec{x}_k + \mat{B} \vec{u}_k + \rvec{w}_k \ , \label{eq:systemX} \\
	\vec{y}_{k}^i  &= \mat{C}^i \rvec{x}_k + \rvec{v}_k^i \ , \label{eq:systemY}
\end{align}
where $\rvec{x}_k \in \mathbb{R}^n$, $\vec{u}_k \in \mathbb{R}^m$, and $\vec{y}_k^i \in \mathbb{R}^{q_i}$ denote the plant's state, the control input applied by the actuator and the measured output of the i-th sensor (with $i \in \mathbb{N}_{>0}$ and $i \leq M$). The matrix \mbox{$\mat{A} \in \mathbb{R}^{n \times n}$} is assumed to be regular. The matrices $\mat{A}$, \mbox{$\mat{B} \in \mathbb{R}^{n \times m}$}, and \mbox{$\mat{C}^i \in \mathbb{R}^{q_i \times n}$} are known by the controller and the sensors. The terms $\rvec{w}_k \in \mathbb{R}^n$ and $\rvec{v}_k^i \in \mathbb{R}^{q_i}$ represent mutually independent, zero-mean, Gaussian white noise processes with covariance matrices $\mat{\Xi} \in \mathbb{R}^{n \times n}$ and $\mat{\Theta}^i \in \mathbb{R}^{q_i \times q_i}$. The initial state of the plant is assumed to be Gaussian distributed with mean $\overline{\vec{x}}_0$ and covariance matrix $\overline{\mat{P}}_0$.

%
%
The digital network connections are subject to time-varying transmission delays and stochastic packet losses. The stochastic characteristics of these effects are assumed to be known. Furthermore, a TCP-like protocol is used for transmissions in the CA-channel. This means that successfully transmitted data packets to the actuator are acknowledged at the controller within the same time step\footnote{A TCP-like network does not reflect real Ethernet-TCP/IP networks, where acknowledgments may be subject to a considerable time delay. However, it is possible to realize a TCP-like network connection by, e.g., message prioritization. Furthermore, TCP-like networks are also of theoretical interest since they give insights into the far more complex cases of real TCP connections or connections where no acknowledgments are provided by the network (as, e.g., for Ethernet UDP/IP networks).}. For transmissions between sensors and controller we relax this assumption and consider the less restrictive case that no acknowledgments are available. Finally, we assume the network nodes to be synchronized, data packets to be tagged with a time-stamp, and controller, actuator, and sensors to be time-triggered.

%
%
To compensate for time delays and data losses between controller and actuator, at every time step the controller sends control sequences to the actuator that not only contain a control input intended for application in the current time step $\vec{u}_{k|k}$, but also $N_A \in \mathbb{N}_{0}$ control inputs for consecutive future time instants. Such a control packet is referred to as $\vec{U}_k$ and is of the form
\begin{align}
	\vec{U}_k = \begin{bmatrix}\vec{u}_{k|k}\tr & \vec{u}_{k+1|k}\tr & \dots & \vec{u}_{k+N_A|k}\tr \end{bmatrix}\tr \ , \label{eq:sequenceEntries}
\end{align}
where, for example, $\vec{u}_{k+1|k}$ denotes a control input calculated at time step $k$ that is intended to be applied at time step $k+1$. The actuator is equipped with a buffer to store the control sequence with the most recent information (among all received sequences). By applying the control input of this buffered sequence that matches the current time step, time delays and losses of subsequent data packets can be compensated until a packet with more recent information arrives. In case the actuator does not receive a new control sequence until the buffer runs out of applicable control inputs, the actuator applies a known default control input $\vec{u}^d$.
		
%
%
As a counterpart to the CA-channel, network effects between sensors and controller are also compensated by a sequence-based design philosophy. More precisely, the sensors not only use the measurement of the current time step $k$ to generate a data packet but also include measurements of the last $N_S \in \mathbb{N}_0$ time steps. The information used to generate the output packet of the i-th sensor $\mathcal{Z}_k^{{\rm out}, i}$ can formally be described by
\begin{align}
	\mathcal{Z}_k^{{\rm out,}i} = g(y_{k-N_S:k}^i) \ , \label{eq:sensorAlgo}
\end{align}
where $g(\cdot)$ denotes an arbitrary sensor algorithm and the notation $y_{a:b}$ denotes the set of measurements $\{y_k | a \leq k \leq b\}$.
The union of all $M$ sets $\mathcal{Z}_k^{{\rm out,}i}$ is denoted by
\begin{align}
	\mathcal{Z}_k^{\rm out} = \mathcal{Z}_k^{{\rm out}, 1} \cup \mathcal{Z}_k^{{\rm out}, 2} \cup \cdots\cup \mathcal{Z}_k^{{\rm out}, M} \ .
\end{align}
Due to network disturbances, the controller may receive no, one, or even more than one packet per sensor at each time step. The set of packets received by the controller of the i-th sensor at time step $k$ is defined by the set $\mathcal{Z}_k^{{\rm in}, i}$ and the union of these sets at time step $k$ over all $M$ sensors by $\mathcal{Z}_k^{{\rm in}}$.

%
%
The problem investigated in this paper is to find an admissible control law that minimizes the quadratic cost function
\begin{align}
	C_0^K &= \Ew \left\{ \rvec{x}_K\tr \mat{Q} \rvec{x}_K  + \displaystyle\sum_{k=0}^{K-1} \rvec{x}_k\tr \mat{Q} \rvec{x}_k + \vec{u}_k\tr \mat{R} \vec{u}_k \Bigg{|} \vec{U}_{0:K-1}, \overline{\vec{x}}_0, \overline{\mat{P}}_0\right\}	\ ,  \label{eq:Cost02k} 
\end{align}
where $K \in \mathbb{N}_{>0}$ is the final time step and $\mat{Q}$ and $\mat{R}$ are symmetric weighting matrices that are positive semi-definite and positive definite, respectively. A control law is called admissible if it requires only information that is available to the controller. In our case, the information available to the controller is described by the information set
\begin{align}
	\mathcal{I}_k = \left\{ \overline{\vec{x}}_0, \overline{\mat{P}}_0, \mathcal{Z}^{\rm in}_{1:k}, \vec{U}_{0:k-1}, \theta_{0:k-1} \right\} \ \label{eq:Ik} ,
\end{align}
where $\theta_k$ represents the information provided by the acknowledgments of the TCP-like connection between controller and actuator.

\section{Optimal Control Law}
\label{sec:Optimal_Controller_Design}
%
%
To solve the sequence-based optimal control problem formulated in Sec.~\ref{sec:Problem_Formulation}, we use a result derived in~\cite{FischerArxiveACC12}, where a simpler problem setup was considered containing only one sensor that transmits one raw measurement per time step. It turns out that the results can easily be extended to the considered setup with multiple sensors that send (possibly pre-processed) measurement information. We summarize the extended result in the following theorem that is formulated in terms of the augmented state 
\begin{align}
	\rvec{\xi}_k &=
	\begin{pmatrix} 
		\rvec{x}_k \\
		[\vec{u}_{k|k-1}\tr \ \ \vec{u}_{k+1|k-1}\tr \ \ \cdots \ \ \vec{u}_{k+N-1|k-1}\tr ]\tr \\
		[\vec{u}_{k|k-2}\tr \ \ \vec{u}_{k+1|k-2}\tr \ \ \cdots \ \ \vec{u}_{k+N-2|k-2}\tr]\tr \\
		\ \ \ \ \vdots\\
		\ \ \ \ \ \ [\vec{u}_{k|k-N+1}\tr \ \ \vec{u}_{k+1|k-N+1}\tr]\tr \\
		\ \ \ \ \ \vec{u}_{k|k-N} \\
		\ \ \ \ \ \vec{u}^d
	\end{pmatrix},
	\label{eq:eta} 
\end{align}
that contains the state of the plant and all control inputs of already sent control sequences that still could be applied to the plant. The last entry is the default control input.
\begin{Theorem}
	\label{theo:sep}
	Consider the problem of finding an optimal control law to generate control sequences of length \mbox{$N_A \in \mathbb{N}_{>0}$} minimizing the cost function~\eqref{eq:Cost02k}	subject to the available information~\eqref{eq:Ik}, system dynamics~\eqref{eq:systemX}, actuator logic described in Sec.~\ref{sec:Problem_Formulation}, and $M$ sensors that process data according to~\eqref{eq:systemY} and \eqref{eq:sensorAlgo}. Then,
\begin{enumerate}
			\item similar as in standard LQG control, the separation principle holds, i.e., the optimal control law can be separated into a) an optimal state estimator that calculates the conditional expectation $\Ew \left\{ \rvec{\xi}_k | \mathcal{I}_k \right\}$ and b) an optimal state feedback controller that utilizes the feedback matrix $\mat{L}_k$,
			
			\item the optimal control law is linear in the conditional expectation of the augmented state according to $$\vec{U}_k = \mat{L}_k \cdot \Ew \left\{ \rvec{\xi}_k  | \mathcal{I}_k \right\} \ ,$$
			
			\item and the state feedback matrix $\mat{L}_k$ explicitly depends on the delay probability distribution of the controller-actuator-network. It can be calculated as given in Theorem 1 of~\cite{FischerArxiveACC12} since it is identical to the feedback matrix for the same system with only one sensor sending one raw measurement per time step.
	\end{enumerate}
\end{Theorem}
\begin{Proof}
			When we replace the measurement equation and information set $\mathcal{I}_k$ of~\cite{FischerArxiveACC12} with the corresponding equations \eqref{eq:systemY} and \eqref{eq:Ik} of our problem, Lemma 1 of~\cite{FischerArxiveACC12} still holds. Therefore, the conditional expectation $\Ew \left\{ \rvec{\xi}_k | \mathcal{I}_k \right\}$ is stochastically independent of the control sequence $\vec{U}_k$. This implies that the separation principle holds and thus, proves part one of the theorem. Part two and three can be proved analogously to the proof of Theorem 1 in~\cite{FischerArxiveACC12} with $\mathcal{I}_k$ replaced by \eqref{eq:Ik}. 
\end{Proof}

\begin{Remark}
It is interesting to note that Theorem~\ref{theo:sep} also holds for all information structures that can be characterized by cumulative subsets $\widetilde{\mathcal{I}}_k$ of the information set $\mathcal{I}_k$, i.e., by subsets that satisfy $\widetilde{\mathcal{I}}_{k-1} \subseteq \widetilde{\mathcal{I}}_k$ and $\widetilde{\mathcal{I}}_k \subseteq \mathcal{I}_k$. This holds regardless of the choice of $N_S$, i.e., how much information of previous time steps is transmitted per packet. Therefore, Theorem~\ref{theo:sep} also holds for the case of infinite measurement sequences.
\end{Remark}

%
%
According to Theorem~\ref{theo:sep}, the optimal control law consists of the combination of an optimal state estimator and an optimal state feedback controller. Since the optimal state feedback controller is the same as in~\cite{FischerArxiveACC12}, we focus on finding the optimal sequence-based state estimator to calculate the conditional expectation ${\rm E} \left\{ \rvec{\xi}_k | \mathcal{I}_k \right\}$ in the following. 

\section{Optimal Estimator}
\label{sec:Estimator_Design}
%
%
As discussed in the introduction, one way to implement a sequence-based information approach for the SE-channel is to include the last $N_S$ measurements into the sensor output packet at every time step. This corresponds to setting $g \equiv \text{id}$ in~\eqref{eq:sensorAlgo}, and results in
\begin{align}
	\mathcal{Z}_k^{{\rm out}, i} = \left\{ y_k^i, y_{k-1}^i, \cdots, y_{k-N_S}^i \right\} 
\end{align}
for the output packet of the i-th sensor with $y^i_r = \emptyset$ for $\ \ r < 1$. Let $\mathcal{Z}_{1:k}^{\rm in}(N_S)$ denote the set of all received measurement sets $\mathcal{Z}^{\rm out, i}_t$ with $1 \leq t \leq k$, $i \in \{1,\dots,M\}$ at the controller side conditioned on $N_S$ and $\mathcal{I}_k (N_S)$ the corresponding information set. Then, 
\begin{align}
	\mathcal{Z}_{1:k}^{\rm in}(0) \subseteq \mathcal{Z}_{1:k}^{\rm in}(1) \subseteq \dots \subseteq \mathcal{Z}_{1:k}^{\rm in}(k) \subseteq \mathcal{I}_k 
\end{align}
and thus, the estimation accuracy of the conditional mean 
\begin{align*}
{\rm E} \left\{ \rvec{\xi}_k \Big{|} \mathcal{I}_k \right\} = {\rm E} \left\{ \rvec{\xi}_k \Big{|} \mathcal{I}_k  (N_S) \right\} 
\end{align*}
is best when the complete sequence of measurements is transmitted to the controller, i.e., $N_S = k$. As this is not realizable due to the growing size of the data packet, we seek to find a recursive estimation algorithm that gives the same accuracy as ${\rm E} \left\{ \rvec{\xi}_k \Big{|} \mathcal{I}_k(k)\right\}$, i.e., an algorithm that provides the same results as a central Kalman filter. Note, that in the given scenario, ${\rm E} \left\{ \rvec{\xi}_k \Big{|} \mathcal{I}_k(k)\right\}$ is identical to ${\rm E} \left\{ \rvec{x}_k \Big{|}  \mathcal{I}_k(k)\right\}$ as the $u$'s in \eqref{eq:eta} are known. 

In the following, an extension of the so called Hypothesizing Distributed Kalman Filter~(HKF) is derived that calculates ${\rm E} \left\{ \rvec{x}_k \Big{|} \mathcal{I}_k(k)\right\}$ based on locally pre-processed measurements when an assumption about the global measurement model has been met by the estimates. 

%
%
\subsection{Hypothesizing Distributed Kalman Filter}
The idea of the HKF is to process local estimates in a transformed state space and filter measurements according to gains that are optimized according to a global measurement model. As the globalization of gains leads to biased local estimates in general, a correction matrix is maintained that allows to eliminate the induced bias. 

A detailed derivation of the algorithm including consistent mean-squared-error (MSE) matrix bounds is given in~\cite{Reinhardt2012b}. More precisely, the correctness of a central version of the HKF has been derived in~\cite{Reinhardt2012a} and has been extended in~\cite{Reinhardt2012b} to the concept of maintaining local variables that allow the calculation of a combined correction matrix. 
In this paper we limit ourselves to the presentation of key formulas and derive the subsequent incorporation of control inputs in more detail afterwards. All derivations and conclusions are given for the time-invariant system from Sec.~\ref{sec:Problem_Formulation}, but are applicable one-to-one to the time-variant case.

%
%
Let $G^f_k \subseteq \{1,\dots, M\}$ contain indices of sensors whose estimates are available at the controller. In order to apply the HKF, a \textbf{h}ypothesis about the \textbf{g}lobal \textbf{m}easurement \textbf{m}odel~(HGMM), i.e., 
\begin{align}
\label{eq:hkf_assumption}
\inv{ \infCz{k} }  \approx \sum_{i \in G^f_k} \transpose{ \infH{i}{k} } \inv{\infCv{i}{k}}  \infH{i}{k}
\end{align}
is necessary, which is chosen according to experimental data or is updated iteratively. Although an unbiased estimate is provided by the HKF in any case, the estimation quality in terms of the MSE matrix depends on the choice of the HGMM and is only equivalent to the globally optimal one when the HGMM meets the sum of the measurement models in~\eqref{eq:hkf_assumption}.

%
%
We consider sensors that have no prior information and that initialize local values $\infx{i}{k}$ and $\infCx{i}{k}$ by help of the first locally obtained measurement according to
\begin{align}
\label{eq:hkf_init}
\infx{i}{1} = \infL{i}{1} \truez{i}{1} \text{ and } \infCx{i}{1} = \infCz{1}
 \text{ with } \infL{i}{k} = \infCx{i}{k}  \transpose{ \infH{i}{k} } \inv{\infCv{i}{k}} \text{ .}
\end{align}
When initial estimates $\estx{i}{0}$, $\estCx{i}{0}$ are given on sensor side, we set
\begin{align*}
\infx{i}{0} = \infCx{i}{0} ( \estCx{i}{0} )^{-1} \estx{i}{0} \text{ and } \infCx{i}{0} = \bigg( \sum_{i=0}^N ( \estCx{i}{0} )^{-1} \bigg)^{-1} \text{ .}
\end{align*}
%
%
The variables are predicted according to
\begin{align*}
\infx{i}{k+1} = \infA{k}{} \infx{i}{k} \text{ and } \infCx{i}{k+1} = \infA{k}{} \infCx{i}{k} \transpose{\infA{k}{}}  + \infCw{k} \text{ .}
\end{align*}
Using $\infL{i}{k}$ from~\eqref{eq:hkf_init} and $\infK{k} = \infCx{i}{k} \inv{\infCx{i}{k|k-1} }$, the filter operation is given by
\begin{align*}
\infx{i}{k} \! = \! \infK{k} \infx{i}{k|k-1}  \! + \!  \infL{i}{k} \truez{i}{k} \text{ and }  
\infCx{i}{k} \! = \! \big( ( \infCx{i}{k|k-1} )^{-1} \!\! + \! ( \infCz{k} )^{-1} \big)^{-1} \!\!  \text{ .}
\end{align*}
For further processing, it is sufficient to transmit $\infx{i}{k}$ to the controller, where the fused estimate 
\begin{align}
\label{eq:hkf_fusion}
\infx{f}{k} = \sum_{i \in G^f_k} \infx{i}{k}
\end{align}
is obtained. It has been shown that $\infx{f}{k}$ equals the (central) linear minimum MSE result when equation~\eqref{eq:hkf_assumption} holds exactly with $G^f_k$ representing the sources of the estimates that are available~\cite{Reinhardt2012a}. Otherwise, $\infx{f}{k}$ is biased and must be corrected.

%
%
To this end, a further variable $\infDelta{i}{k}$ is maintained that allows to reconstruct unbiased estimates independently of $G^f_k$ and of the HGMM. The idea is to make sure that $\E{ \truex{k} } = \E{ ( \infDelta{i}{k} )^{-1} \infx{i}{k} }$ holds for all local estimates at all time steps. Thus, we initialize the correction matrix with
\begin{math}
\infDelta{i}{1} = \infL{i}{1} \infH{i}{1}
\end{math}
when the estimates are initialized from measurements, or~-- when initial estimates are utilized~-- with
\begin{math}
\infDelta{i}{1} = \infCx{}{1} ( \estCx{i}{1} )^{-1} \text{ .}
\end{math}
The prediction of the correction matrix is led back to the last time step by
\begin{align*}
\infDelta{i}{k+1} = \infA{k}{} \infDelta{i}{k} \inv{ \infA{k}{} } \text{ ,}
\end{align*}
and in the filter step, we set
\begin{align*}
\infDelta{i}{k} = \infK{k} \infDelta{i}{k|k-1} + \infL{i}{k} \infH{i}{k} \text{ .}
\end{align*}
The correction matrix that corresponds to the fused estimate $\infx{f}{k}$ is given by
\begin{align}
\label{eq:hkf_fusion_delta}
\infDelta{f}{k} = \sum_{i \in G^f_k} \infDelta{i}{k} \text{ .}
\end{align}

%
%
It is worth mentioning that $\infDelta{i}{1}$ is not regular at the initialization step when the rank of $\infH{i}{1}$ is below the state dimension. However, this problem also occurs in the central processing schema and is likewise solved after fusing values that together cover the complete state. 
Apart from that, measurement models do not need to be known at remote sensors as potential differences between the HGMM and the actually utilized models in~\eqref{eq:hkf_assumption} can be corrected by helps of $\infDelta{}{k}$.

%
%
\subsection{Subsequent Inclusion of Control Inputs}
Up to now, we have not considered control inputs from the state space model~\eqref{eq:systemX}, which complicates the estimation process as the local sensors have to recursively estimate the state without having information about the deterministic inputs and thus, without knowing the complete system model. In the following, we determine the part of the control inputs that was not comprised in the processed measurements and needs to be added to the estimate. In a second step, we prove that the derived procedure provides globally optimal estimates iff the basic HKF is optimal.

%
%
In order to simplify the calculations, we w.l.o.g. expect prediction and filter steps to be alternating. This assumption allows us to find sum formulas for the key variables. We define 
\begin{align*}
\infG{l\cdot\cdot k}{l\cdot\cdot k} = \bigg( \prod_{t=l}^{k-1} \transpose{\infA{t}{}} \transpose{\infK{t+1}}  \bigg)^\top
\text{ and }
\inv{\infA{l\cdot\cdot k}{l\cdot\cdot k}} = \prod_{t=l}^k \inv{\infA{t}{}} \text{ ,}
\end{align*}
with $\inv{ \infA{t\cdot\cdot l}{t\cdot\cdot l}} = \mat{I}$ when $t=l$ or $t=l+1$, and $\inv{ \infA{t\cdot\cdot l}{t\cdot\cdot l}} := \infA{l+1\cdot\cdot t-1}{l+1\cdot\cdot t-1}$ when $t > l + 1$. Hence, the local estimation values and correction matrices are obtained as
\begin{align*}
\infx{i}{k} \!  =  \!  \sum_{t=1}^k \infG{t\cdot\cdot k}{t\cdot\cdot k} \infL{i}{t} \truez{i}{t} \text{ and }
\infDelta{i}{k} \! = \!  \sum_{t=1}^k \! \infG{t\cdot\cdot k}{t\cdot\cdot k} \infL{i}{t} \infH{i}{t} \inv{\infA{t\cdot\cdot k-1}{t\cdot\cdot k-1}} \!\! \text{ .}
\end{align*}
The values of the fused estimate result in
\begin{align}
\label{eq:hkf_x_recursive} 
\infx{f}{k} = \sum_{t=1}^k \infG{t\cdot\cdot k}{t\cdot\cdot k} \Big( \sum_{i \in G^f_k} \infL{i}{t} \truez{i}{t} \Big) 
\end{align}
and
\begin{align}
\label{eq:hkf_delta_recursive}
\infDelta{f}{k} =  \sum_{t=1}^k \infG{t\cdot\cdot k}{t\cdot\cdot k} \Big( \sum_{i \in G^f_k} \infL{i}{t} \infH{i}{t} \Big) \inv{\infA{t\cdot\cdot k-1}{t\cdot\cdot k-1}} \text{ ,} 
\end{align}
which verifies the key result from~\cite{Reinhardt2012b} that the estimation result of the HKF does not depend on whether information are processed locally or centrally as long as they are finally combined. 
The true state is given by
\begin{align}
\label{eq:hkf_x_true_recursive}
\truex{k} = \sum_{t=0}^{k-1} \Big( \infA{t+1\cdot\cdot k-1}{t+1\cdot\cdot k-1} \infB{t} \trueu{t} + \truew{t} \Big) + \infA{0\cdot\cdot k-1}{0\cdot\cdot k-1} \truex{0}
\end{align}
and, hence, measurements can be represented as
\begin{align}
\label{eq:hkf_z_recursive}
\truez{i}{t} = \infH{i}{t} \bigg( \sum_{l=0}^{t-1} \Big( \infA{l+1\cdot\cdot t-1}{l+1\cdot\cdot t-1} \infB{l} \trueu{l} + \truew{l} \Big) + \infA{0\cdot\cdot t-1}{0\cdot\cdot t-1} \truex{0} \bigg) + \! \truev{i}{t} \! \text{ .}
\end{align}
By means of these formulas, we derive the part of the control inputs that is already included in $\infx{f}{k}$ and thus, determine the vector that still needs to be added in order to yield an unbiased estimate. We obtain:
%
%
\begin{Theorem}
\label{hkf_theorem_1}
The variable 
\begin{math}
\estx{f}{k} =  \inv{\infDelta{f}{k}} \big( \infx{f}{k} + \infxu{f}{k} \big)
\end{math}
with
\begin{align}
\label{eq:hkf_theorem_u1}
\infxu{f}{k} = \sum_{t=0}^{k-1} \infG{t\cdot\cdot k}{t\cdot\cdot k} \infDelta{f}{t} \inv{\infA{t}{}} \infB{t} \trueu{t}
\end{align}
is an unbiased estimate of a system with dynamics~\eqref{eq:systemX}.
\end{Theorem}
\begin{Proof}
\ref{sec:hkf_theorem_1}.
\end{Proof}

%
%
With $\infxu{f}{k}$ from Theorem~\ref{hkf_theorem_1}, we are able to reconstruct an unbiased estimate at the controller even if measurements are processed locally at the sensors. But, as we do not expect to have the same estimates available at every time step, we need the following corollary that allows to maintain the control input parts for every node separately.

%
%
\begin{Corollary}
The variable $\infxu{f}{k}$ from~\eqref{eq:hkf_theorem_u1} equals the sum of node-specific variables
\begin{align}
\label{eq:hkf_corollary_1}
\infxu{i}{k} = \sum_{t=0}^{k-1} \infG{t\cdot\cdot k}{t\cdot\cdot k} \infDelta{i}{t} \inv{\infA{t}{}} \infB{t} \trueu{t} \text{ .}
\end{align}
\end{Corollary}
\begin{Proof}
Simple matrix algebra using~\eqref{eq:hkf_delta_recursive}.
\end{Proof}

%
%
It is worth mentioning that~\eqref{eq:hkf_corollary_1} depends on the control input as well as on the actually utilized model at the node (implicitly by $\infDelta{i}{t}$) and thus, both types of information have to be available at either the controller or the sensors. In this paper, we focus on the first case, where $\infDelta{i}{t}$ is obtained at the controller. Nevertheless, it would also be possible to communicate the control inputs directly from the actuator(s) to the sensors and transmit $\infxu{i}{k}$ in combination with $\infx{i}{k}$ to the controller. 

%
%
Before we finish the derivations, we provide a Lemma concerning the optimality of the proposed procedure.
\begin{Lemma}
The HKF with subsequent control input inclusion is globally optimal when the HGMM has met the sum of the actually utilized measurement models from~\eqref{eq:hkf_assumption} at all time steps.
\end{Lemma}
\begin{Proof}
When the global measurement model is met, $\infDelta{f}{t} = \mat{I}$, $\forall t \in {1,\dots,k}$ holds, and therefore, the fused estimate $\estx{f}{k}$ is obtained by
\begin{align*}
\inv{\infDelta{f}{k}} \big( \infx{f}{k} + \infxu{f}{k} \big) =  \infx{f}{k} + \infxu{f}{k} \stackrel{ \eqref{eq:hkf_x_recursive} \eqref{eq:hkf_theorem_u1}}{=} 
\sum_{t=1}^k \infG{t\cdot\cdot k}{t\cdot\cdot k} \Big( \sum_{i \in G^f_k} \infL{i}{t} \truez{i}{t} \Big) +
\sum_{t=0}^{k-1} \infG{t\cdot\cdot k}{t\cdot\cdot k} \inv{\infA{t}{}} \infB{t} \trueu{t}  = \\
\sum_{t=1}^{k} \infG{t\cdot\cdot k}{t\cdot\cdot k} \Big( \sum_{i \in G^f_k} \infL{i}{t} \truez{i}{t}  + \infK{t} \infB{t-1} \trueu{t-1} \Big) \text{ .}
\end{align*}
By splitting up $\infK{t}$ and $\infL{i}{t}$, we obtain
\begin{align*}
\sum_{t=1}^{k} \infG{t\cdot\cdot k}{t\cdot\cdot k} \infCx{i}{t} \Big( \! \sum_{i \in G^f_k} \transpose{ \infH{i}{t} } \inv{\infCv{i}{t}} \truez{i}{t} \!  + \! \inv{\infCx{}{t|t-1}} \infB{t-1} \trueu{t-1} \Big) \text{ ,}
\end{align*}
which equals the recursive information form of the Kalman Filter when all measurements are processed centrally and therefore, is the linear optimal solution.
\end{Proof}

\subsection{Application}
In the following, the derived approach is applied to the plant that has been depicted in Sec.~\ref{sec:Problem_Formulation}. In order to minimize the communication and computation at the sensors, we calculate only $\infx{i}{k}$ at the distributed nodes and maintain $\infxu{i}{k}$ and $\infDelta{i}{k}$ at the controller. As estimates can be delayed or lost, the controller stores the measurement models and control inputs since the last time step of successfully received variables $\infx{i}{t}$, $t < k$. By helps of this information, the estimation variables are predicted according to 
\begin{align*}
\infxBackup{i}{k} = \infG{t\cdot\cdot k}{t\cdot\cdot k} \infx{i}{t} \text{  , } 
\infDeltaBackup{i}{k} = \infG{t\cdot\cdot k}{t\cdot\cdot k} \infDelta{i}{t} \inv{ \infA{t\cdot\cdot k-1}{t\cdot\cdot k-1} } 
\end{align*} 
and
\begin{align*}
\infxuBackup{i}{k} = \infxu{i}{t} + \sum_{l=t}^{k-1} \infG{l\cdot\cdot k}{l\cdot\cdot k} \infDelta{i}{l} \inv{\infA{l}{}} \infB{l} \trueu{l}  =
\infxu{i}{t} +  \sum_{l=t}^{k-1} \! \infG{t\cdot\cdot k}{t\cdot\cdot k}   \infDelta{i}{t}  \! \inv{ \infA{t\cdot\cdot l}{t\cdot\cdot l} } \infB{l} \trueu{l} 
=  \\
\infxu{i}{t} +  \infG{t\cdot\cdot k}{t\cdot\cdot k} \infDelta{i}{t}  \sum_{l=t}^{k-1}  \inv{ \infA{t\cdot\cdot l}{t\cdot\cdot l} } \infB{l} \trueu{l} 
\end{align*}
when no estimate of the current time step is available. If it is likely that transmissions fail, it is meaningful to store these predicted variables as future predictions are based on them. Otherwise, it is sufficient to hold variables $\infx{i}{t},\infDelta{i}{t},\infxu{i}{t},\infCx{i}{t}$ for the latest estimate for each sensor, and $\trueu{k}$ for each time step for which there is a chance that an estimate is predicted from. 

%
%
Although several matrices have to be stored and calculated at the controller, this is unlikely to be a problem in real-world applications where it is necessary to spatially separate controller, actuators, and sensors at the costs of a poor communication. Apart from that, optimal out-of-sequence algorithms, which are the alternative to the proposed algorithm, have at least comparable costs.

%
%
It remains to discuss how the HGMM is chosen best. In order to apply the HKF it is possible to set the HGMM to an arbitrary matrix. However, as we are interested in optimizing the estimation quality at the controller, two strategies are meaningful in the context of the presented scenario. When the HGMM is set to the sum of the measurement models of all available nodes, i.e., $G^f_k = \{1,\dots,M\}$ in~\eqref{eq:hkf_assumption}, the estimate at the controller is globally optimal when all estimates from one time step are received. This is especially meaningful when packet losses and delays are unlikely. Alternatively, the HGMM is chosen according to the expected ``measurement quality'', e.g., as the half of the sum of all measurement models when the sensor measurement models are similar and it is expected that slightly more than half of the transmissions fail. 

%
%
In plants with slowly changing communication quality, a more sophisticated approach can be employed. Although the exact procedure is out of the scope of this paper, it is reasonable to adapt the HGMM when the fused correction matrix~\eqref{eq:hkf_fusion_delta} differs substantially from the identity matrix over multiple time steps. In this case, the HGMM should be reduced in dimensions with negative deviation to the identity matrix and increased otherwise. 

%
%
Independent of the specific strategy, the estimation result is almost optimal when the HGMM approximately meets the actually utilized models. In~\cite{Reinhardt2012c}, it has been shown for multiple systems and measurement models in a sensor network consisting of ten nodes that the MSE of the HKF does not exceed the one of the linear optimal solution by more than six percent when the actually utilized global measurement model from~\eqref{eq:hkf_assumption} does not deviate by more than 40 percent from the HGMM. 

%
%
In summary, we have proven that by means of $\infxu{i}{k}$ from~\eqref{eq:hkf_corollary_1}, it is possible to reconstruct an unbiased estimate from recursively gained information that is globally optimal when the HGMM fits to the measurement models of the available estimates, even if the control inputs are not known to the sensors but are subsequently included at the controller. In order to apply the proposed algorithm, we have discussed strategies to maintain the relevant variables and to choose the HGMM appropriately. 

\section{Conclusion}
\label{Conclusion}
%
%
In this contribution, a unified approach to sequence-based control for networked control systems was proposed. We have extended recent results on the optimal sequence-based control for systems with only one sensor sending raw data to the case of multiple sensors, that process measurement data in a sequence-based information framework. To this end, we introduced and extended the Hypothesizing Distributed Kalman Filter as distributed estimation algorithm. Although we have not presented an estimator that is equivalent to the central Kalman Filter in all cases, we have suggested methods for choosing a meaningful hypothesis about the global measurement model so that the estimator provides slightly suboptimal to optimal estimates when the communication model does not change abruptly.

%
%
Future research will focus on relaxing the TCP-like assumption on the controller-actuator-channel and on suitable approximations of the multiple actuator scenario. The estimation procedure can be improved by handling measurement failures. Apart from that, we see great potential in applying the proposed algorithm to a real plant.

\begin{appendix}
\gdef\thesection{Appendix \Alph{section}}

\section{Proof of Theorem~\ref{hkf_theorem_1}}
\label{sec:hkf_theorem_1}
\begin{Proof}
In order to prove the theorem it is sufficient to show the equality between $\E{ \infx{f}{k} + \infxu{f}{k} }$ and $\E{ \infDelta{f}{k} \truex{k}}$.
First, we note that the inner terms of $\infxu{f}{k}$ from~\eqref{eq:hkf_delta_recursive} are given by
\begin{align}
\label{eq:hkf_theorem_1_proof_u}
\infG{l\cdot\cdot k}{l\cdot\cdot k} \infDelta{f}{l} \inv{\infA{t}{}}  \stackrel{ \eqref{eq:hkf_delta_recursive} }{=} \sum_{t=1}^l \infG{t\cdot\cdot k}{t\cdot\cdot k} \Big( \sum_{i \in G^f_k} \infL{i}{t} \infH{i}{t} \Big) \inv{\infA{t\cdot\cdot l}{t\cdot\cdot l}} \text{ .}
\end{align}
The expected value of the estimate without considering control inputs is given by
\begin{align*}
\E{ \infx{f}{k} } \stackrel{ \eqref{eq:hkf_x_recursive} }{=} \sum_{t=1}^{k} \infG{t\cdot\cdot k}{t\cdot\cdot k} \Big( \sum_{i \in G^f_k} \infL{i}{t} \E{ \truez{i}{t} } \Big) \stackrel{ \eqref{eq:hkf_z_recursive} }{ =} 
\sum_{t=1}^{k} \infG{t\cdot\cdot k}{t\cdot\cdot k} \!\!  \sum_{i \in G^f_k}  \! \infL{i}{t} \infH{i}{t} \bigg( \sum_{l=0}^{t-1} \! \infA{l+1\cdot\cdot t-1}{l+1\cdot\cdot t-1} \infB{l} \trueu{l} + \infA{0\cdot\cdot t-1}{0\cdot\cdot t-1} \E{ \truex{0} } \!\!  \bigg) \! \text{ ,}
\end{align*}
which is transformed in order to factorize the control inputs to
\begin{align}
\label{eq:hkf_theorem_proof_1} 
\sum_{l=0}^{k-1}  \bigg( \sum_{t=l+1}^k \infG{t\cdot\cdot k}{t\cdot\cdot k}  \Big( \sum_{i \in G^f_k} \infL{i}{t} \infH{i}{t}  \Big) \infA{l+1\cdot\cdot t-1}{l+1\cdot\cdot t-1} \bigg) \infB{l} \trueu{l} + 
\sum_{t=1}^{k} \infG{t\cdot\cdot k}{t\cdot\cdot k} \sum_{i \in G^f_k} \infL{i}{t} \infH{i}{t} \infA{0\cdot\cdot t-1}{0\cdot\cdot t-1} \E{ \truex{0} }  \text{ .}
\end{align}
The inner term of~\eqref{eq:hkf_theorem_proof_1} can be combined with the corresponding inner term of $\infxu{f}{k}$ that is given by $\infG{l\cdot\cdot k}{l\cdot\cdot k} \infDelta{f}{l} \inv{\infA{t}{}}$ as
\begin{align*}
\sum_{t=l+1}^k \infG{t\cdot\cdot k}{t\cdot\cdot k} \Big( \sum_{i \in G^f_k} \infL{i}{t} \infH{i}{t}  \Big)  \infA{l+1\cdot\cdot t-1}{l+1\cdot\cdot t-1} + \infG{l\cdot\cdot k}{l+1\cdot\cdot k} \infDelta{f}{l} \inv{\infA{}{}}
 \stackrel{ \eqref{eq:hkf_theorem_1_proof_u} }{=}  \sum_{t=1}^k \infG{t\cdot\cdot k}{t\cdot\cdot k} \Big( \sum_{i \in G^f_k} \infL{i}{t} \infH{i}{t}  \Big) \inv{ \infA{t\cdot\cdot l}{t\cdot\cdot l}} \text{ ,}
\end{align*}
and thus, we obtain for $\E{ \infx{f}{k} + \infxu{f}{k} }$
\begin{align}
\sum_{l=0}^{k-1} \bigg( \sum_{t=1}^k \infG{t\cdot\cdot k}{t\cdot\cdot k} \Big( \sum_{i \in G^f_k} \infL{i}{t} \infH{i}{t}  \Big) \inv{ \infA{t\cdot\cdot l}{t\cdot\cdot l}} \bigg) \infB{l} \trueu{l} + 
 \sum_{t=1}^{k} \infG{t\cdot\cdot k}{t\cdot\cdot k} \sum_{i \in G^f_k} \infL{i}{t} \infH{i}{t} \infA{0\cdot\cdot t-1}{0\cdot\cdot t-1} \E{ \truex{0} } \label{eq:hkf_theorem_sum_1}  \text{ .}
\end{align}
With~\eqref{eq:hkf_delta_recursive} and~\eqref{eq:hkf_x_true_recursive}, the second term $\E{ \infDelta{f}{k} \truex{k}}$ is given by
\begin{align*}
\bigg( \sum_{t=1}^k \infG{t\cdot\cdot k}{t\cdot\cdot k} \Big( \sum_{i \in G^f_k} \infL{i}{t} \infH{i}{t} \Big) \inv{\infA{t\cdot\cdot k-1}{t\cdot\cdot k-1}}  \bigg)
\bigg( \sum_{l=0}^{k-1} \infA{l+1\cdot\cdot k-1}{l+1\cdot\cdot k-1} \infB{l} \trueu{l} + \infA{0\cdot\cdot k-1}{0\cdot\cdot k-1} \E{ \truex{0} } \bigg) \text{ ,}
\end{align*}
which is simplified with $\inv{\infA{t\cdot\cdot k-1}{t\cdot\cdot k-1}} \infA{l+1\cdot\cdot k-1}{l+1\cdot\cdot k-1} = \inv{\infA{t\cdot\cdot l}{t\cdot\cdot l}}$ to~\eqref{eq:hkf_theorem_sum_1}.
\end{Proof}
%
%
\end{appendix}

\section*{Acknowledgments}
This work was partially supported by the German Research Foundation (DFG) within the Research Training Group GRK 1194 ``Self-organizing Sensor-Actuator-Networks'' and within the Priority Program 1305 ``Control Theory of Digitally Networked Dynamical Systems''.

\bibliographystyle{IEEEtran}
\bibliography{Paper}

\end{document}